\documentclass[fleqn,usenatbib, useAMS]{mnras}

\usepackage[T1]{fontenc}
\usepackage{ae,aecompl}
\usepackage[normalem]{ulem}

\usepackage{graphicx}	
\usepackage{amsmath}	
\usepackage{amssymb}	
\usepackage{ragged2e}
\usepackage{newtxtext,newtxmath}
\usepackage[nodayofweek]{datetime}
\newdateformat{monthyearday}{%
  \THEYEAR\ \monthname[\THEMONTH] \twodigit{\THEDAY}}

\makeatletter
\patchcmd{\NAT@citex}
  {\@citea\NAT@hyper@{%
     \NAT@nmfmt{\NAT@nm}%
     \hyper@natlinkbreak{\NAT@aysep\NAT@spacechar}{\@citeb\@extra@b@citeb}%
     \NAT@date}}
  {\@citea\NAT@nmfmt{\NAT@nm}%
   \NAT@aysep\NAT@spacechar\NAT@hyper@{\NAT@date}}{}{}

\patchcmd{\NAT@citex}
  {\@citea\NAT@hyper@{%
     \NAT@nmfmt{\NAT@nm}%
     \hyper@natlinkbreak{\NAT@spacechar\NAT@@open\if*#1*\else#1\NAT@spacechar\fi}%
       {\@citeb\@extra@b@citeb}%
     \NAT@date}}
  {\@citea\NAT@nmfmt{\NAT@nm}%
   \NAT@spacechar\NAT@@open\if*#1*\else#1\NAT@spacechar\fi\NAT@hyper@{\NAT@date}}
  {}{}
\makeatother

\title[Observations of Three Young M-Stars in Upper Sco]{An HST/STIS View of Protoplanetary Disks in Upper Scorpius: Observations of Three Young M-Stars}

\author[S. Walker et al.]{
Sam Walker,$^{1}$\thanks{E-mail: s1602779@ed.ac.uk}
Maxwell Andrew Millar-Blanchaer,$^{2,3}$
Bin Ren,$^{3}$
Paul Kalas$^{4,5,6}$ and
\newauthor ~John Carpenter$^{7}$\\
$^{1}$SUPA\thanks{Scottish Universities Physics Alliance}, Institute for Astronomy, University of Edinburgh, Royal Observatory, Edinburgh EH9 3HJ, UK\\
$^{2}$Department of Physics, University of California, Santa Barbara, CA 93106, USA\\
$^{3}$Department of Astronomy, California Institute of Technology, 1216 East California Boulevard, Pasadena, CA 91125, USA\\
$^{4}$Astronomy Department, University of California, Berkeley, CA 94720, USA\\
$^{5}$SETI Institute, Carl Sagan Center, 189 Bernardo Ave.,  Mountain View CA 94043, USA\\
$^{6}$Institute of Astrophysics, FORTH, GR-71110 Heraklion, Greece\\
$^{7}$Joint ALMA Observatory, Avenida Alonso de C\'ordova 3107, Vitacura, Santiago, Chile}

\date{Accepted XXX. Received YYY; in original form \monthyearday\today}

\pubyear{2020}

\begin{document}
\label{firstpage}
\pagerange{\pageref{firstpage}--\pageref{lastpage}}
\maketitle

\begin{abstract}
We present observations of three protoplanetary disks in visible scattered light around M-type stars in the Upper Scorpius OB association using the STIS instrument on the Hubble Space Telescope. The disks around stars 2MASS J16090075--1908526, 2MASS J16142029--1906481 and 2MASS J16123916--1859284 have all been previously detected with ALMA, and 2MASS J16123916--1859284 has never previously been imaged at scattered light wavelengths. We process our images using Reference Differential Imaging, comparing and contrasting three reduction techniques -- classical subtraction, Karhunen-Lo\`eve Image Projection and Non-Negative Matrix Factorisation, selecting the classical method as the most reliable of the three for our observations. Of the three disks, two are tentatively detected (2MASS J16142029--1906481 and 2MASS J16123916--1859284), with the third going undetected. Our two detections are shown to be consistent when varying the reference star or reduction method used, and both detections exhibit structure out to projected distances of ${\gtrsim}200~\textrm{au}$. Structures at these distances from the host star have never been previously detected at any wavelength for either disk, illustrating the utility of visible-wavelength observations in probing the distribution of small dust grains at large angular separations.


\end{abstract}



\section{Introduction}

Protoplanetary disks provide a window into the development of all types of planetary systems \citep{andrews2020}. These disks are both precursors to and indicators of planet formation, and give insight into the very early stages of the life of a planetary system. It is in these environments that the dust and gas that constitute a typical protoplanetary disk are able to coalesce to form the vast array of planets known to science \citep{winn2015}, from small rocky worlds like Earth \citep[e.g.][]{fressin2012, jenkins2015, kaltenegger2019}, to gas giants with orbital periods from years \citep[e.g.][]{gaudi2008, blunt2019, feng2019} down to only a few days \citep[e.g.][]{mayor1995, yu2017, zhou2019}. While advances have been made in understanding the dynamics of protoplanetary disks and how these contribute to forming the planetary zoo, many questions remain unanswered. There is vast scope to increase understanding of the demographic properties of the protoplanetary disk population, and how this varies as a function of host star spectral type, stellar age, and stellar multiplicity; another issue of importance is characterisation of disk substructure to help guide theoretical frameworks of planet formation \citep{andrews2020}.


 Statistical analyses of protoplanetary disk populations suggest that the dispersal timescales of these disks must be rapid (removing the gas on scales of ${\leqslant}10$ Myr; e.g. \citealt{zuckerman1995}) and mass-dependent \citep[although the nature of this dependence varies between studies operating at different wavelength regimes; e.g.][]{carpenter2006, barenfeld2016, pascucci2016}.
Two central mechanisms are thought to be involved in this dispersal: accretion onto the star \citep{hartmann2016} and photoevaporation, the star-driven heating of the disk to temperatures sufficient to excite disk material out of the gravitational potential well of the star \citep{hollenbach2000}.
Accretion onto the star can occur for both the dust and gas within the disk, whereas most dust grains are too heavy to be directly removed via photoevaporation, although smaller ($\leqslant 100\mu$m) dust grains remain coupled to the gas in the disk and can thus be evacuated along with it \citep[e.g.][]{takeuchi2002}. 
Consequently, the dust-to-gas ratio increasingly favours larger ($\geqslant 1$mm) dust grains in the disk over time \citep{dubrulle1995}, which has been shown to positively affect the creation of planetesimals, and may play a crucial role in the creation of rocky planets and debris disks \citep{throop2005, wyatt2008}.
Further work is needed to extend our understanding of the role that disk dispersal can have in producing substructure and influencing planet formation. Key to this are resolved images of disks, which enable the detection of features such as gaps, rings and spiral arms that may indicate the presence of planets in formation or disk dispersal in action \citep[e.g.][]{avenhaus2014, dong2015, boccaletti2020, ren2020, wang2020}.     

Observations in differing regions of the electromagnetic spectrum can be used in conjunction to form a holistic view of disk evolution, as different wavelengths of light probe distinct elements of a typical protoplanetary disk \citep{sicilia2016}. Observations at millimetre and sub-millimetre wavelengths can be used to detect continuum emission from larger mm-scale dust grains in the disk, as well as spectral line emissions from tracer molecules in the gas. These data can be used to directly probe the temperature of the gas \citep[e.g.][]{heese2017} and the solids surface density \citep[e.g.][]{isella2009, pietu2014} in the disk and indirectly probe the mass of the gas \citep[e.g.][]{andrews2013, ruiz2018}. Visible/near-infrared observations can probe the micron and sub-micron dust distribution via observations of light from the host star scattered by the dust in the disk towards the observer. This provides a valuable tracer of the dust in the surface layers of the disk, which can be especially useful in observing the flaring of a protoplanetary disk \citep[e.g.][]{wolff2017, avenhaus2018}. The surface brightness of a protoplanetary disk in scattered light is dependent on the scattering properties of the dust grains in addition to the disk structure, allowing investigation of the dust itself \citep{sicilia2016}.
   
In this paper we present new scattered light observations of three M-type stars in the Upper Scorpius OB association. This association (hereafter known as Upper Sco) is a star-forming region at a distance of $\sim$145 pc \citep{preibisch2008} that plays host to M-type stars with ages of 5-11 Myr \citep{preibisch2002, pecaut2012}, up to 22\% of which have circumstellar disks \citep{luhman2020}, thus providing a unique opportunity for the study of protoplanetary disks. The ages of these stars are typical of the projected lifetimes of disks around low-mass stars 
\citep{mamajek2009},
and previous studies of the association have detected protoplanetary disks at an advanced stage \citep{scholz2007, luhman2012}, highlighting Upper Sco as an important region for studies of protoplanetary disks near the end of their lifetimes. Smaller M-type stars also allow easier detection of Earth-like planets using either the transit or radial velocity methods compared to larger stars. Therefore, examining protoplanetary disks around M-stars like our targets can give an insight into the formation mechanisms that give rise to the Earth analogues that can be found orbiting such stars. The proximity of Upper Sco to Earth is also ideal for high-resolution direct imaging of protoplanetary disks \citep[e.g.][]{mayama2012, zhang2014, barenfeld2016, dong2017, garufi2020}.
Our work aims to address a gap in the knowledge of disks around M-stars by imaging disks around three targets in scattered light, one for the first time, with a view to using these images to study the evolution of small dust grains around late-type stars.

We discuss the observations of each star in Section \ref{observations}, before detailing the data reduction steps taken to produce our final science images in Section \ref{datareduction}. The results of the analysis of these final science images are presented in Section \ref{results}, followed by a brief conclusion.

\section{Observations}
\label{observations}

\subsection{Target selection}

The Upper Sco region was the focus of a wider survey observing protoplanetary disk $0.88$~mm continuum and $^{12}$CO $J = 3-2$ line fluxes with ALMA as detailed in \citet{barenfeld2016}. The results of these observations were used to derive masses for the dust observed in the disk, and in a follow-up paper surface density profiles were fit to the observations to better understand the morphology of the observed disks \citep{barenfeld2017}. This survey found 6 M-type stars that hosted disks with radii within the \textit{HST}/STIS field of view and greater than the inner working angle of the STIS BAR5 occulter \citep[$0\farcs2$;][]{debes2019} that had not been previously observed in scattered light. The three most massive of these disks with the most favourable inclinations were selected for the observations in scattered light using STIS detailed below.

\subsection{Observations}

We adopt the strategy of Reference Differential Imaging \citep[RDI, ][]{smith1984} to remove the starlight and reveal the surrounding circumstellar structure (see Section~\ref{datareduction} for further details), choosing to observe one reference star per science target. We observed our three targets (2MASS J16090075--1908526, 2MASS J16142029--1906481 and 2MASS J16123916--1859284) and corresponding three reference stars (2MASS J16132214--1924172, 2MASS J16082234--1930052 and 2MASS J16150856--1851009) under \href{https://www.stsci.edu/hst/phase2-public/15176.pdf}{GO 15176} and \href{https://www.stsci.edu/hst/phase2-public/15497.pdf}{GO 15497} (PI: M. Millar-Blanchaer) with \textit{HST}/STIS using the BAR5 occulter. We list the stellar and observation parameters in Table \ref{tab:observations}. The $G_{\textrm{BP}} - G_{\textrm{RP}}$ colour data presented in Table \ref{tab:observations} are obtained from \textit{Gaia} EDR3 \citep{gaiaedr3}. These colours were used as they are readily available for all six stars and cover almost the full range of STIS sensitivity, with the two filters neatly bisecting the STIS sensitivity range. It should be noted, however, that these were not the colours used to select the reference stars, as each reference star was selected to match the $B - V$ colour and $V$-band magnitude of its corresponding target star.

For each science target, exposures were obtained at two telescope roll angles (one orbit per roll), separated by between $19^\circ$ and $27^\circ$. Each target's corresponding reference star was observed for one orbit at one roll angle. Six exposures were taken at each roll angle, for a total of 12 exposures per target star and 6 exposures per reference star. This strategy of observing at multiple roll angles was utilised to average out instrumental variations and ensure that any observed structure was not merely a function of the STIS detector. Our observations were taken with the star positioned not at the default BAR5 location but rather a custom location at the tip of the BAR5 occulter to obtain data at the smallest possible working angle. This involved specifying the target acquisition on BAR10 (another of the STIS occulters) with the positional offset parameter POS TARG set to X = 16\farcs34596, Y = $-7$\farcs17672.

\begin{table*}
\caption{The observation parameters and stellar information for each target and reference star observed under \href{https://www.stsci.edu/hst/phase2-public/15176.pdf}{GO 15176} and \href{https://www.stsci.edu/hst/phase2-public/15497.pdf}{GO 15497}. 
}

\label{tab:observations}
\setlength{\tabcolsep}{2pt}
\begin{tabular}{ccccccccccccc}
\hline
    2MASS ID & Identifier & SpT & $G_{\textrm{BP}}{-}G_{\textrm{RP}}$ &Distance&$R_{\textrm{dust}}$ & $\theta_{\textrm{dust}}$ & $i_{\textrm{dust}}$ &$R_{\textrm{CO}}$ & $\theta_{\textrm{CO}}$& $i_{\textrm{CO}}$& Exposures & Roll angle(s)\\
    & & & &(au) & (au) & ($^{\circ}$) & ($^{\circ}$) & (au) & ($^{\circ}$) & ($^{\circ}$) & & ($^{\circ}$) \\
    (1) & (2) & (3) & (4) & (5) & (6) & (7) & (8) & (9) & (10) & (11) & (12) &(13) \\
\hline
J16090075--1908526 & Target 1    & K9 & $2.19\pm0.03$ & $137.4\pm0.5$ &$58^{+5}_{-4}$  & $149^{+9}_{-9}$  & $56^{+5}_{-5}$ & $169^{+24}_{-26}$ & $104^{+14}_{-11}$ & $53^{+6}_{-8}$ & 379s${\times}12$ & $65.1$, $92.1$
\\
J16132214--1924172 & Reference 1 & K3 & $1.491\pm0.005$ & && & & &&& 379s${\times}6$& $62.3$
\\ \hline
J16142029--1906481 & Target 2    & M0 & $2.4\pm0.1$ & $140\pm1$&$29^{+1}_{-2}$ & $19^{+32}_{-19}$ & $27^{+10}_{-23}$ & $88^{+6}_{-6}$ & $5^{+4}_{-4}$ & $58^{+4}_{-4}$ & 379s${\times}12$ & $-124.9$, $-104.9$
\\
J16082234--1930052 & Reference 2 & M1 & $2.41\pm0.01$ & & & &&&&&379s${\times}6$ & $-124.2$
\\ \hline
J16123916--1859284 & Target 3    & M0.5 & $2.33\pm0.03$ & $135.5\pm0.3$ &$48^{+8}_{-7}$ & $46^{+22}_{-27}$ & $51^{+14}_{-36}$ & $72^{+42}_{-22}$ & $95^{+62}_{-40}$ & $50^{+10}_{-41}$ &369s${\times}12$ & $74.1$, $90.1$
\\
J16150856--1851009 & Reference 3 & M0 & $2.48\pm0.01$ & &&&&& & &369s${\times}6$ & $63.2$
\\ \hline
\end{tabular}
\justify
{\textbf{Notes}: (1) 2 Micron All Sky Survey (2MASS) identifier. (2) Reference $x$ is the star observed contemporaneously with Target $x$, and was selected to provide the best match for that target's PSF. (3) The spectral type of the star as obtained from \citet{luhman2012}, with the exception of Reference 1, whose spectral type is inferred from its effective temperature as reported in \citet{rave2016}. (4) The colour of the star as obtained from \citet{gaiaedr3}. (5) The distance to the star from the parallaxes presented in \textit{Gaia} EDR3. (6-11) Disk information obtained from the fits to ALMA $0.88$~mm continuum and $^{12}$CO $J = 3-2$ observations as detailed in \citet{barenfeld2017} -- disk radius $R$ is the maximum disk radius such that the surface density profile of the disk is 0 for $r > R$, $\theta$ is the position angle of the semi-major axis of the disk, and $i$ the inclination measured from face-on. (12) Number of STIS exposures, and length of each exposure. (13) Roll angles used for each STIS orbit, rounded to the nearest $0\fdg1$.}
\end{table*}

There were multiple \textit{HST} gyro failures during our observing period. Observations of J16142029--1906481 and the corresponding reference were carried out on 24/2/2018 before these set of failures, when gyros \#1, \#2 and \#4 were in use. Gyro \#1 then failed and \#6 took over, which led to a much larger jitter (${\sim}16$~mas compared to ${\sim}3$~mas before the failure) and increased RMS noise by a factor of ${\sim}2$ within $0\farcs4$ of the star \citep{stisreport1}. 
J16090075--1908526 and its reference were observed in these conditions on 18/6/2018. Following these observations, gyro \#2 failed, and the current gyros in use are \#3, \#4 and \#6. The jitter is now at ${\sim}7$~mas and contrast is nominal\footnote{\url{https://www.stsci.edu/contents/news/stis-stans/march-2019-stan}} \citep{stisreport2}. J16123916--1859284 and its reference were observed in this current epoch on 13/6/2019.

\section{Data Reduction}
\label{datareduction}

One vital processing step when dealing with scattered light observations of circumstellar disks is the removal or subtraction of the stellar point spread function \citep[PSF; e.g.][]{schneider2014}. In the visible, the observed flux from the star is orders of magnitude larger than that of the scattered light from the surrounding disk. Diffraction and instrumental effects conspire to spread the light from the star across the detector, obscuring any fainter objects beneath. The intensity of the PSF can be suppressed by using a coronagraph, but even when using a coronagraph it is necessary to use innovative post-processing techniques to remove the PSF from observed images \citep[e.g.][]{lafreniere2007, soummer2012, ren2018, pairet2020}.

\subsection{Data Preparation}
Before performing any stellar PSF subtraction, the jitter of the telescope between exposures had to be corrected for. In order to do this, the absolute centre of each star in each frame was located with the \texttt{centerRadon} Python package \citep{centerradon}, which utilises Radon Transforms to perform line integrals along different azimuthal angles for each on-sky location, thus designating the stellar centre as the location that has the maximum line integral value \citep{pueyo2015}. The frames were then aligned to place the centre of the star at the centre of each frame using \texttt{scipy.ndimage.shift}, before the frames were cropped to an area of $120 \times 120$ pixels (${\sim}6\arcsec \times 6\arcsec$ using the STIS platescale of 1 pixel width = 0\farcs05078; \citealt{stisbook2019}) around the stellar centre. The data were then converted from units of counts to those of surface brightness using the conversion equation in Appendix B.2.1 of \citet{viana2009} to convert to flux, and then dividing by the area of sky that one pixel covers to obtain the data in surface brightness units.

To clean the data of bad pixels, we follow \citet{ren2017} in applying a $3 \times 3$ pixel median filter to correct for pixels that were identified in the STIS data quality file extensions either as having a dark rate ${>}5\sigma$ above the median dark level, as being a known bad pixel or as being affected by cosmic rays. In our data reduction, we used a version of the software mask created by \citet{debes2017} to exclude the regions occulted by BAR5, which we increased in size to compensate for some telescope jitter that could not be entirely corrected using our centering algorithm. Smaller masks were experimented with, but were found to produce unreliable results with very large positive and negative residuals at smaller working angles. We also excluded a $9$ pixel wide region centred at the diffraction spikes to minimize the impact of the spikes during data reduction -- the data excluded by this mask can be seen enclosed by the black lines on the example raw image in Figure \ref{fig:mask}.

\subsection{Data Post-processing}
\subsubsection{Methods}
Three methods were used to remove the stellar PSFs from our observations -- a classical reduction (i.e., scaling and subtracting a reference image directly from the science image), Karhunen-Lo\`eve Image Projection \citep[KLIP,][]{soummer2012} and Non-negative Matrix Factorisation \citep[NMF,][]{ren2018}. KLIP is a relatively well-established method of PSF removal that was developed as an iteration on previous, overly aggressive PSF subtraction algorithms, and has been widely used to recover images of directly imaged disks \citep[e.g.][]{soummer2014, mazoyer2016, chen2020}.
However, each target and reference image must be standardised (i.e. target $t \longrightarrow \frac{t - \mu_t}{\sigma_t}$) in order to calculate the covariance matrix as part of the KLIP method. The irreversible loss of flux that results from this standardisation means that forward modelling is often necessitated to fully characterise observed astrophysical features \citep[e.g.][]{pueyo2016, arriaga2020}.
NMF was thus developed as an alternative, and as it involves no such reduction in flux it has been shown to retrieve fainter morphological features without the need for forward modelling for STIS images \citep{ren2018}. A central focus of this study is to compare and contrast these two methods of reduction with the classical method and with each other.

\begin{figure}
    \centering
    \includegraphics[keepaspectratio, width=\columnwidth]{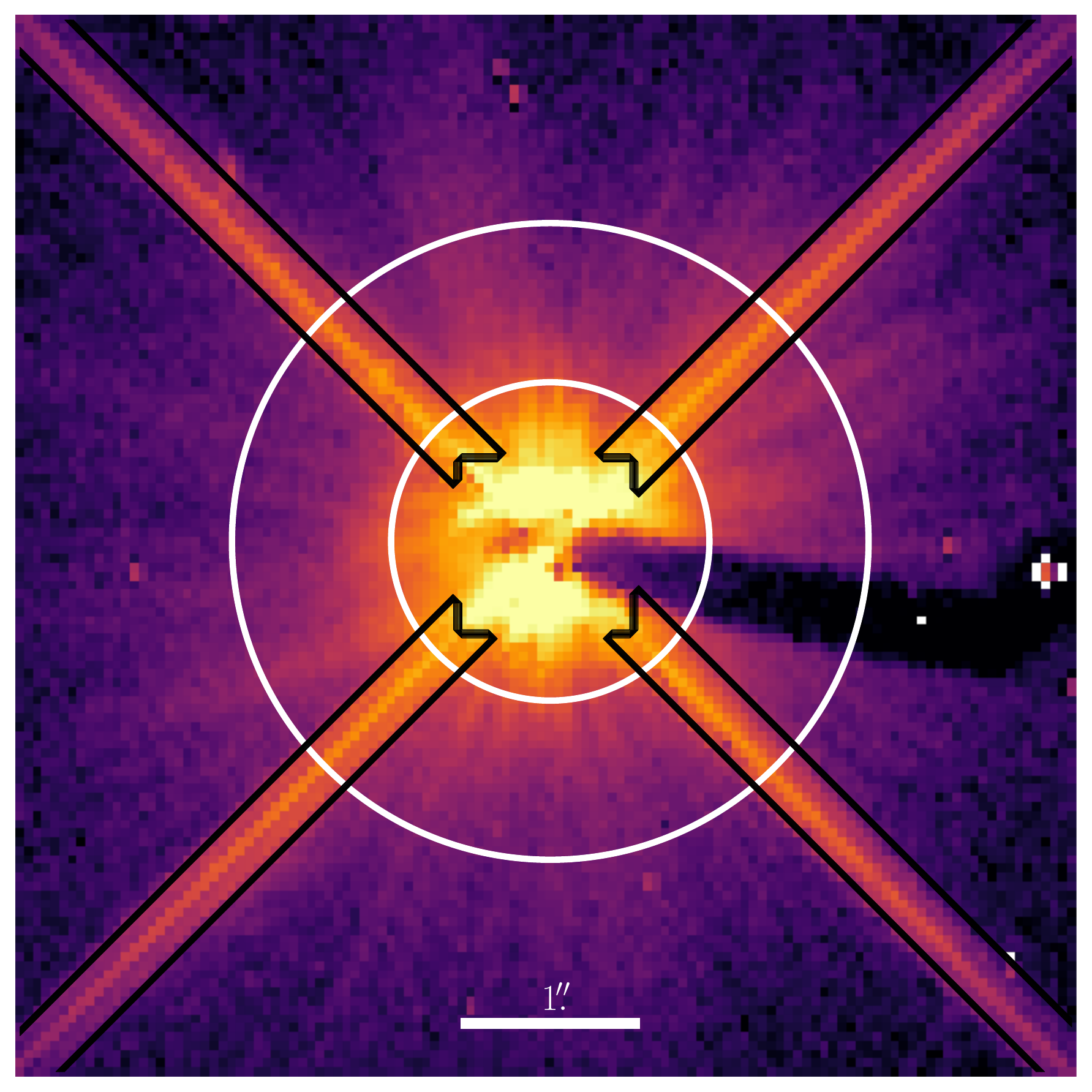}
    \caption{Example raw target image, with the region in which the standard deviation is minimised for the classical method outlined in black. The area over which the azimuthal profiles displayed in Figure \ref{fig:angleprof} were computed is displayed between the two white annuli of radii 0\farcs9 and 1\farcs8. Image is displayed in units of log counts, and the white scalebar indicates 1\farcs~Our custom target acquisition places stars near the tip of the BAR5 occulting element.}
    \label{fig:mask}
\end{figure}

Following the data preparation described above, a library of 18 reference PSFs was constructed using each of the 6 frames from each of our 3 reference stars. Any given PSF subtraction method is highly sensitive to the exact reference frames used as part of the subtraction process. As such, we experimented with two of the three methods of reference frame selection utilised in \citet{ruane2019} -- the Pearson correlation coefficient (PCC) and the structural similarity index matrix \citep[SSIM,][]{ssim}. As described in \citet{ruane2019}, the PCC is designed to select for structural differences between two images, whereas the SSIM is sensitive to differences in brightness as well as in structure. The third method mentioned in \citet{ruane2019} was the mean square error (MSE), but this method involves no image standardisation and appears therefore to be useful primarily for comparing stars with similar fluxes, which is not necessarily the case for our target-reference pairs -- as such we saw no use in applying it here. Our final choice was to use the SSIM, as we find that it better discriminates between references than the PCC, in agreement with \citet{ruane2019}.

\subsubsection{Reduction}

The first method utilised for PSF subtraction was the classical method: for each of the 12 target exposures, the single best-matching reference exposure from the pool of 18 reference PSFs was selected using the SSIM metric. Classical subtraction was then performed using the equation $s = t - f \cdot r,$ where $t$ is the target frame, $r$ the reference frame, $s$ the final science frame and $f$ is a multiplicative factor that minimises the standard deviation in the outer diffraction spikes of $s$. This factor $f$ was determined using the Nelder-Mead algorithm \citep{neldermead} as implemented in \texttt{scipy.optimize.minimize}, and the region used to determine $f$ can be seen outlined in black in Figure \ref{fig:mask}.
After this process was followed for all target exposures, the resulting science frames were rotated such that North was up and East was left. The diffraction spikes and STIS coronagraph regions were then masked, and the median of these frames was taken as the final science image.

KLIP was carried out as per \citet{soummer2012} on each target exposure individually, before rotating and combining as with the classical method above. This method was trialled first without frame selection, allowing the components to be ranked by their eigenvalues, thereby selecting the best $n$ components that are representative of the signals in the reference PSFs. We also employed explicit frame selection to select the best reference PSFs, using all $n$ components generated by the selected $n$ reference PSFs. During each of these processes, the combined BAR5 and diffraction spike mask described above was used to prevent values covered by this mask being used in any of our calculations. Each target and reference frame was standardised by subtracting off the mean and dividing by the standard deviation of the frame before KLIP was performed, in accordance with the methodology. The resulting final science frame was multiplied by the standard deviation of the initial target image to scale it back to its original units. 

The NMF method was implemented using the \texttt{nmf\_imaging} package \citep{nmfpackage}. Similarly to KLIP, NMF was also tested with and without explicit frame selection, using the combined BAR5 and diffraction spike mask at all stages. An option available in the \texttt{nmf\_imaging} package is to calculate and multiply by the multiplicative factor that minimises the standard deviation in the residual image (similar to the factor $f$ described in the classical method, but minimising the standard deviation in all unmasked regions instead of only the outer diffraction spikes).
This was experimented with by implementing separate reductions with and without this factor -- it did not appear to much impact the resulting image, but our final reductions utilised it as recommended in \citet{ren2018}.



We experimented with varying the number of KLIP and NMF modes and found that, in both cases, 5 modes gave a good balance between PSF subtraction and low oversubtraction. Increasing the number of modes did not significantly change the final PSF-subtracted images.


As previously mentioned, one factor that can strongly impact the fidelity of the PSF subtraction is the exact reference PSFs used when performing the subtraction. This was seen distinctly when using Reference 1, as every reduction that used any of the 6 PSFs generated from this reference star was left with significant PSF residuals which were not apparent in reductions solely using reference PSFs obtained from References 2 \& 3. It was therefore decided that the reference PSFs constructed from observations of Reference 1 were unreliable, and as such these were left out of all further analysis. After removing Reference 1's reference PSFs from our library, the resulting images were much more stable when frame selection was varied, increasing confidence that Reference 1 was indeed a poor match to our science targets. There are at least two potential reasons why Reference 1 performed so poorly -- the first of these is that there was a considerable colour mismatch between Reference 1 and our other stars across the full STIS sensitivity range. Even though Reference 1 appeared to be close to our target stars in $B - V$ colour, the mismatch at other wavelengths evidenced by the \textit{Gaia} colours reported in Table \ref{tab:observations} might well have caused the star to perform poorly overall. It could also be that the increased telescope jitter during the period when Reference 1 was observed (see Section \ref{observations}) might have made observations during this time period less directly comparable to observations with lower telescope jitter. However, reductions of J16090075--1908526 (which was observed during the same time period as Reference 1) using only reference PSFs from Reference 1 produced significantly worse results than when using frames from the non-contemporaneous References 2 and 3, which decreases the likelihood of this hypothesis, as the telescope jitter should be comparable for this target/reference pairing. It should also be noted that frames from Reference 1 performed equally well as those from References 2 and 3 with both the PCC and SSIM frame selection indicators, highlighting the necessity of finding more effective and accurate methods of frame selection. Regardless of the root cause, these findings underline the need for very careful reference star selection when reducing scattered light disk images.

\begin{figure*}
\centering
\includegraphics[keepaspectratio, width=\textwidth]{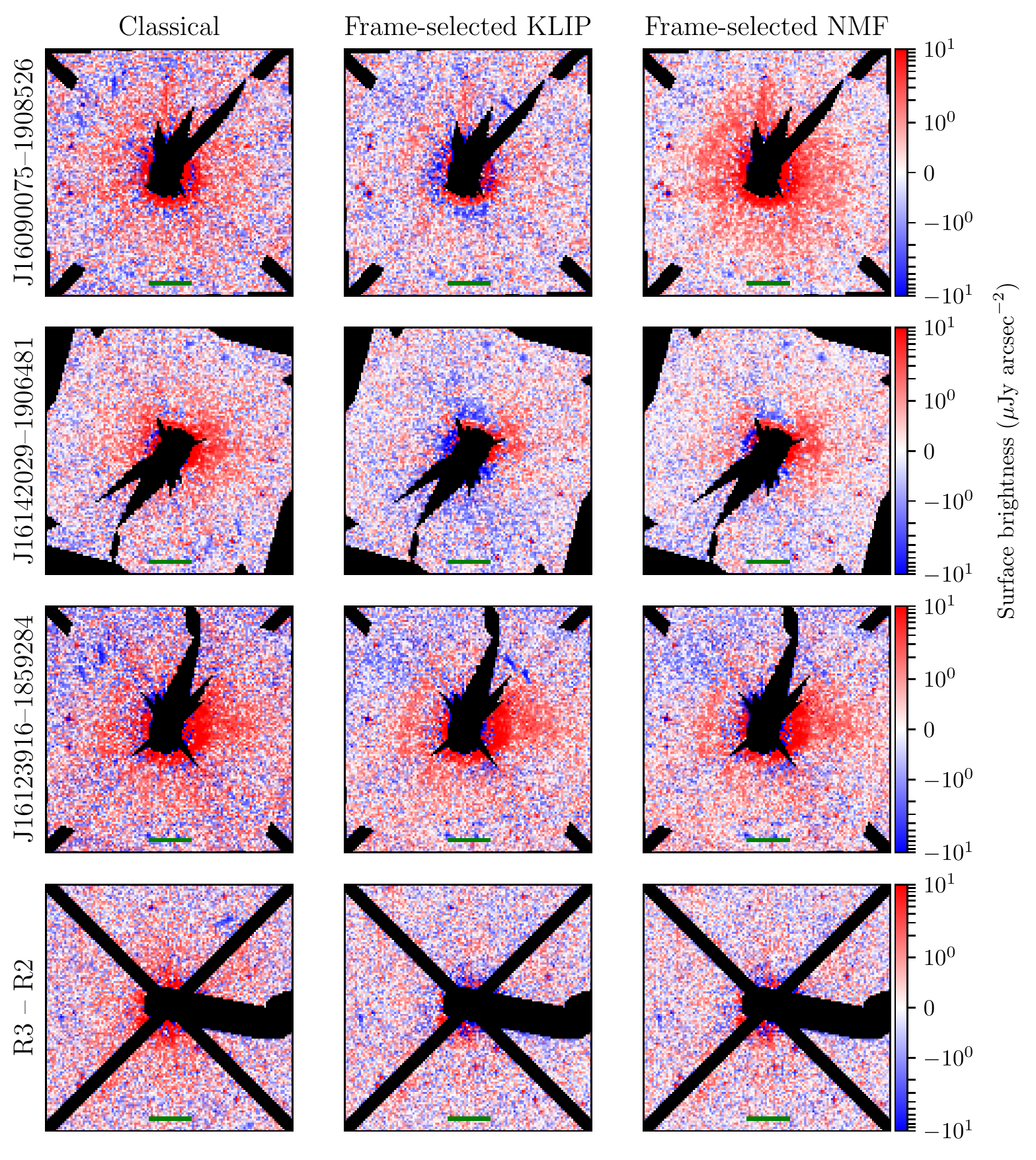}
\caption{The final science images obtained by using each of the labeled methods for each of our three targets. The green scalebars represent 1\arcsec, and the black regions show the areas covered by our combined BAR5 and diffraction spike mask at both roll angles. The final row displays a reduction of Reference 3 using frames from Reference 2, and is included as an example of a null detection. Images are oriented such that up is North and left is East, with the exception of the R3 -- R2 images, which are presented unrotated as they would appear in the detector frame.}
\label{fig:finalplots}
\end{figure*}

To further investigate the effect of reference star choice on the residual image following the discarding of reference PSFs from Reference 1, we expanded our library to include other STIS observations of M-type stars using an updated version of the archive presented in \citet{ren2017}. However, results obtained using this expanded library were of significantly poorer quality than our previous results, with all three target reductions dominated by non-physical artefacts when using any of our methods. In contrast to our images, these library observations were conducted at the default BAR5 location. As such, there are likely differences in flatfielding and PSF behaviour between the two datasets, which could be contributing factors as to why these references performed as poorly as they did. Further to this, the PSF behaviour changes over time due to the `breathing' of the telescope, which could also negatively affect the comparability of the PSF over time \citep{grady2003}. Another such factor could be potential colour mismatches between the expanded library and our three targets, but the fact that these references performed so poorly meant that we did not investigate this further. None of the images in this expanded library were used in our final analysis.


We present in Figure \ref{fig:finalplots} the final science images obtained using each of our three methods, enabling comparisons between them. As expected, the KLIP reduction is the most aggressive, subtracting more flux than the classical and NMF cases and leading to high levels of oversubtraction for both J16090075-–1908526 and J16142029-–1906481. This, coupled with the knowledge that KLIP has been found to eradicate known disk features in STIS images of well-characterised disks \citep{ren2018}, leads us to discount this method for our final analysis. The NMF method seems to be a little less reliable than either of the other two methods for our target stars, as the extended halo present in the NMF result for J16090075--1908526 illustrates -- when performing separate reductions using only frames from Reference 2 and Reference 3 individually, this halo appeared only in reductions using Reference 3 frames. This inconsistency between two reference stars that perform equally well using either KLIP or the classical method leads us to conclude that NMF is less reliable for these observations. As such, we take the classical results to be the most representative of the final disk in each case, plotting them again using an adjusted colour bar and with the median radial profile subtracted to better display the disk structures in Figure \ref{fig:finalfinalplot}. We focus our analysis on these classical PSF-subtracted images.

The main cause of a false positive disk detection is PSF artefacts due to a mismatch between the reference and target PSFs. As mentioned in our discussion of Reference 1 above, the two ways that this might occur are either by temporal PSF variation or colour mismatch between the reference and target stars. The first of these is especially of concern due to the different HST gyro configurations used to observe References 2 \& 3. To investigate whether temporal PSF variation might have been responsible for any structure in Figure \ref{fig:finalfinalplot}, we experimented with separate reductions for each target star using only the 6 frames from each of References 2 and 3 respectively as our PSF reference library. However, these reductions were found to be very similar in each case, increasing confidence that these sets of reference PSFs provide good PSF templates for our target stars. As a further check, we performed the reductions of Reference 3 using Reference 2 shown in the final row of Figure \ref{fig:finalplots}. The clean reduction that results from this illustrates that the two PSFs are a good match for each other, again demonstrating that temporal PSF variation is unlikely to be a factor in producing any observed signal. As for potential colour mismatch, we can see from Table \ref{tab:observations} that all of our stars have reasonably close \textit{Gaia} colours, with J16090075--1908526 being the largest outlier of our stars (with the exception of the unused Reference 1). This relative outlier still produces a very clean reduction (as detailed in Section \ref{subsec:t1} below), showing that the 12 reference PSFs obtained from observations of References 2 \& 3 are good matches for J16090075--1908526's PSF and that colour mismatch is unlikely to be a factor for our other two targets. As both of these two potential root causes can be discounted, we can be cautiously confident that any observed disk signal is real and astrophysical in origin.

\section{Analysis}
\label{results}

\begin{figure*}
\centering
\includegraphics[keepaspectratio, width=\textwidth]{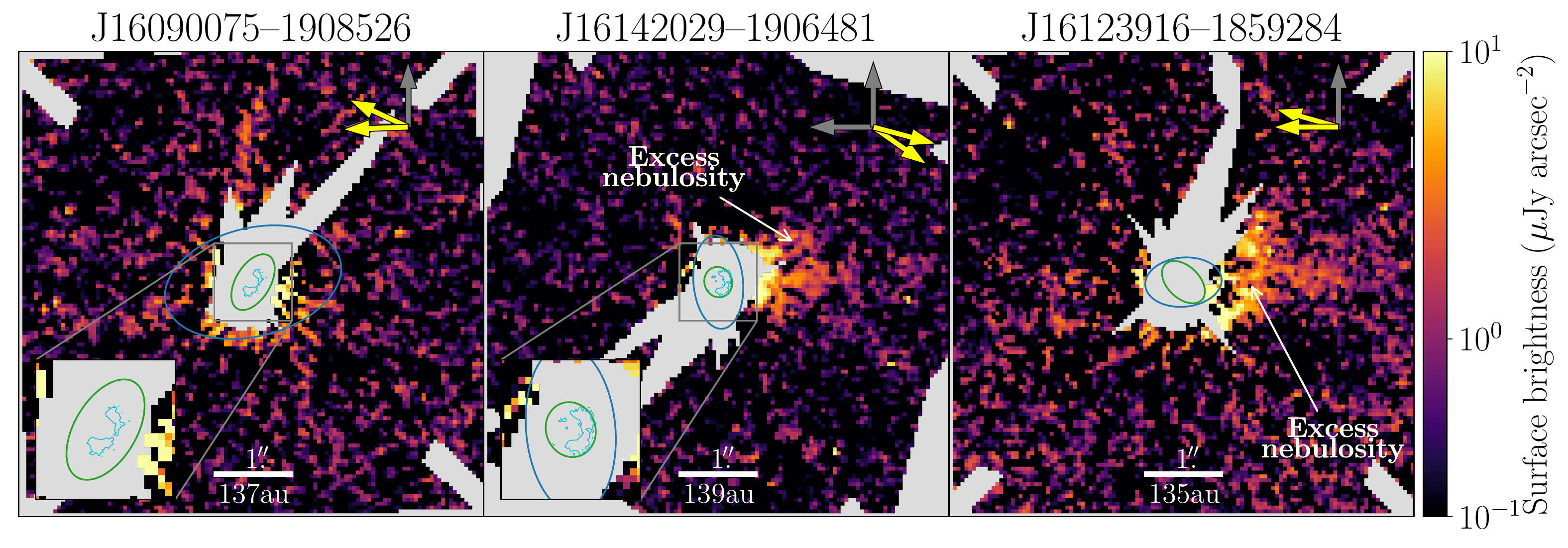}
\caption{The final classical reductions for each of the three targets in our study, with the median radial profile displayed in Figure \ref{fig:radprof} subtracted off and a $3{\times}3$ median filter applied. The dark grey arrows represent North (up) and East (left) for the final images, and the two yellow arrows represent the orientations of North for the initial observations relative to North in the final image. The light grey regions show the areas covered by our combined BAR5 and diffraction spike mask at both roll angles, and the white scalebar represents 1\arcsec and is labelled with the physical length that 1\arcsec corresponds to at the \textit{Gaia} EDR3 distances for each target star. The cyan contours plotted over the mask in the first two images represent (to scale) the corresponding \citet{garufi2020} SPHERE detections of the first two targets, the green and blue ellipses represent the \citet{barenfeld2017} $0.88$~mm continuum dust and $^{12}$CO $J = 3-2$ surface density profiles for each target respectively, again to scale. The regions of excess nebulosities are labelled in the two images that we believe show disk signal.}
\label{fig:finalfinalplot}
\end{figure*}

\begin{figure}
    \centering
    \includegraphics[keepaspectratio, width = \columnwidth]{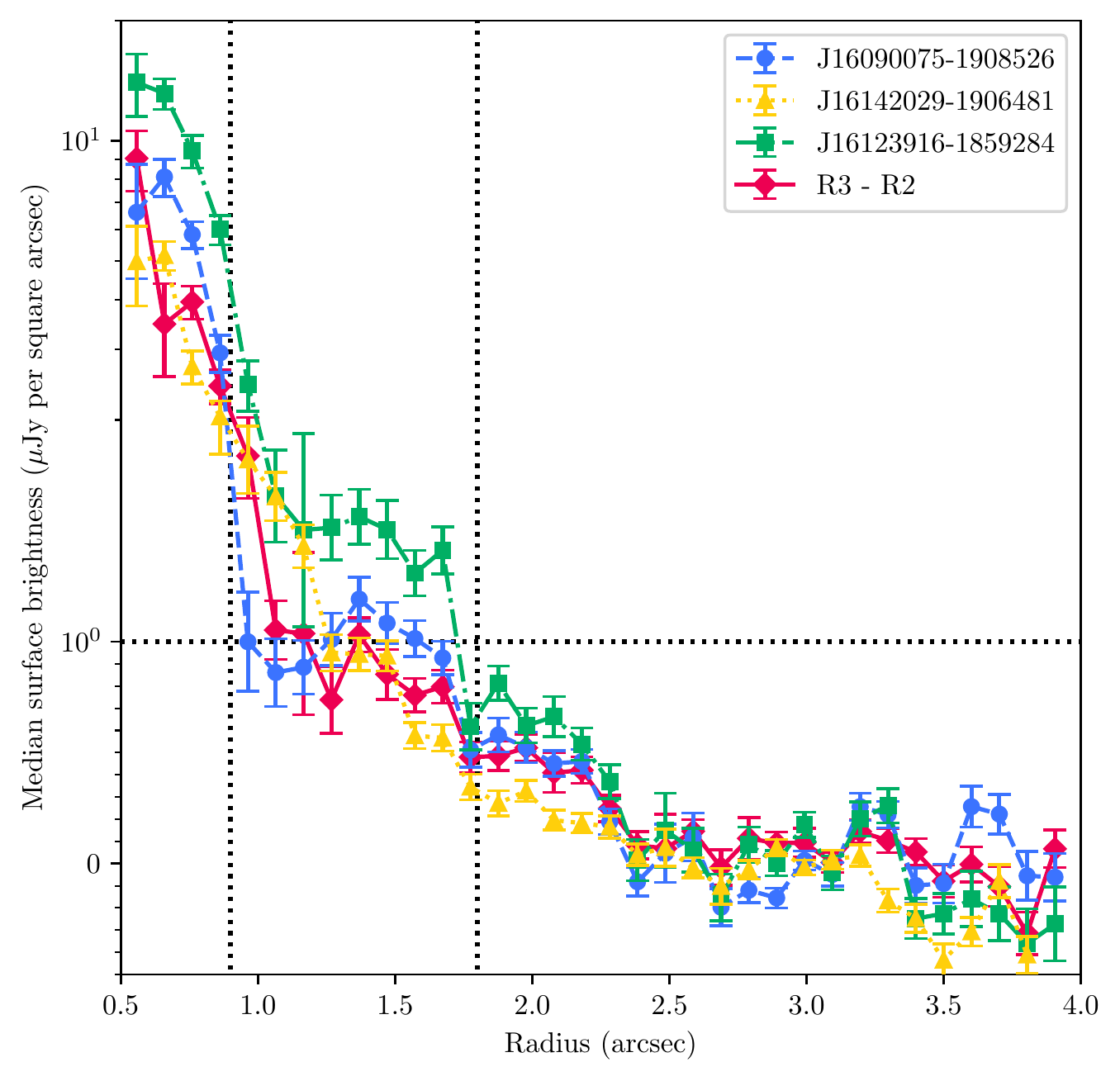}
    \caption{The median surface brightness as a function of radius for the final classical reductions for each target star, as well as for the reduction of reference 3 using reference 2, which serves as a diskless comparison image. These radial profiles were computed as described in Section \ref{results}, with error bars displaying the standard error on each measurement. The dotted black vertical lines indicate the region within which the azimuthal profiles presented in Figure \ref{fig:angleprof} were computed. Note the logarithmic scale for median surface
    brightness $> 1 \mu$Jy per sq. arcsec (above the horizontal black line).}
    \label{fig:radprof}
\end{figure}

\begin{figure}
    \centering
    \includegraphics[keepaspectratio, width=\columnwidth]{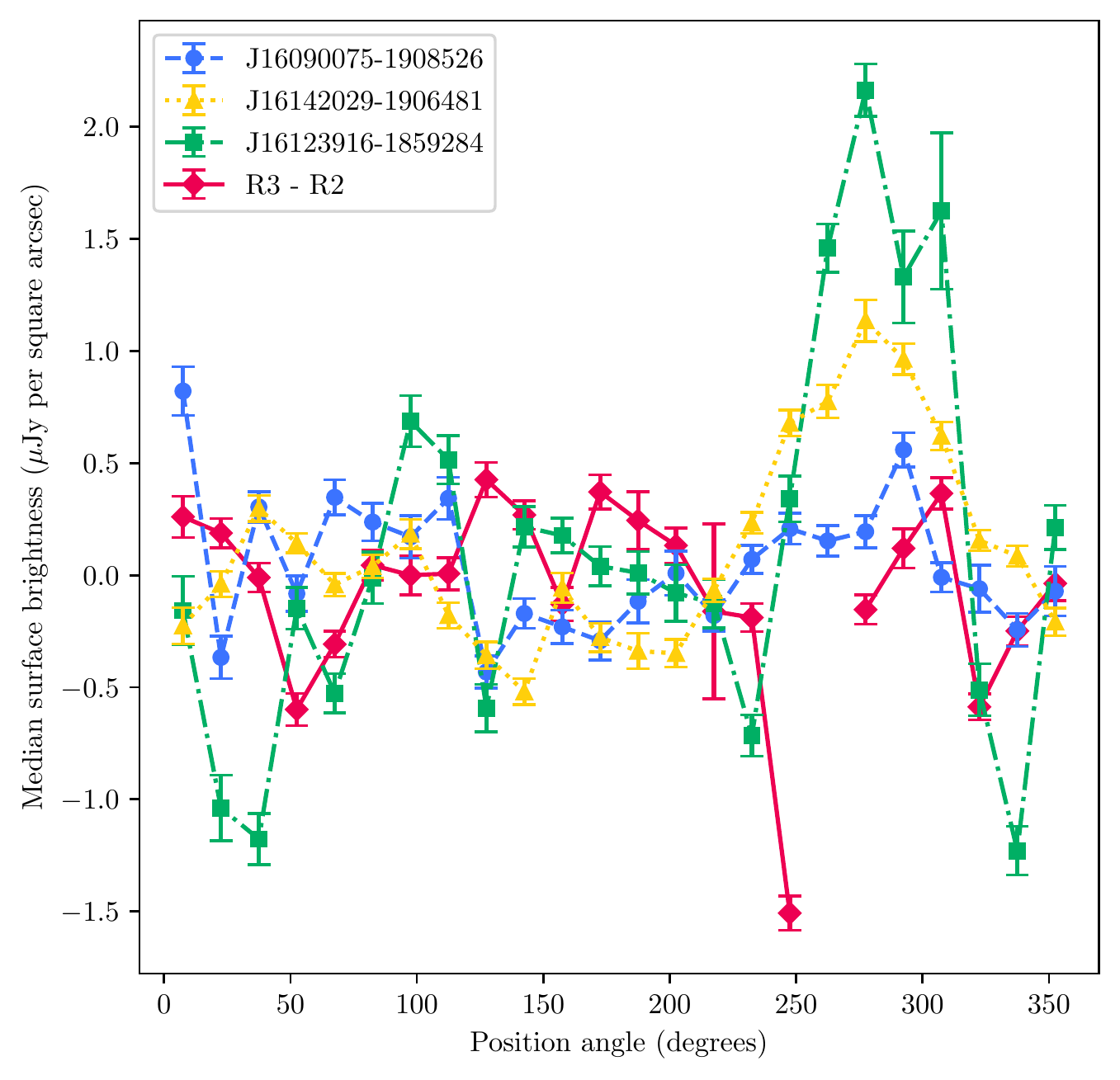}
    \caption{The median surface brightness as a function of on-sky position angle for the final classical reductions for each target star, as well as for the reduction of reference 3 using reference 2, which serves as a diskless comparison image. These azimuthal profiles were computed between angular separations of 0\farcs9 and 1\farcs8 from the star as described in Section \ref{results}, with error bars displaying the standard error on each measurement. For each target, the median surface brightness at all position angles was subtracted off to better highlight the differences in flux with angle.}
    \label{fig:angleprof}
\end{figure}

We characterise the final science images presented in Figure \ref{fig:finalfinalplot} using the radial and azimuthal profiles presented in Figures \ref{fig:radprof} and \ref{fig:angleprof} to understand how the disk signal varies as a function of radius and position angle. The radial profiles were constructed by obtaining the median and standard error from successive 2 pixel wide annuli about the centre of the image, ignoring masked values. The first 3 values for each radial profile were discarded, as the very small numbers of unmasked pixels at such close-in radii led to unreliable samples, leaving us with only those results for separations ${>}0\farcs5$. Different annuli widths from 2 to 5 pixels were experimented with, but these were found to produce similar results, and as an annulus width of 2 pixels already ensures the data are Nyquist-sampled it was felt that there was no need to lose any additional information by increasing the annulus width. The azimuthal profiles were created by computing the median and standard error within $15^\circ$ wide wedges anticlockwise from North within an annulus from 18 to 36 pixels ($0\farcs9$ to $1\farcs8$) in radius about the centre of the image (Figure~\ref{fig:mask} displays these annuli overplotted on an example raw image). Pixels outside this region appear to be almost all noise, as can be seen both by inspection of the science images in Figures~\ref{fig:finalplots} and \ref{fig:finalfinalplot} and from the radial profiles in Figure \ref{fig:radprof}, and pixels within the inner ring of the annulus were discarded due to the relatively small sample sizes at small angular separations, which could otherwise allow several bright pixels to severely bias the overall azimuthal profiles and obscure the trends at moderate separations that much better illustrate whether or not a disk signal is truly present. Wedge sizes of 5, 10, 15 and 30 pixels were experimented with, but were found not to significantly affect the plotted results, hence 15 pixels was chosen as a compromise between high levels of sampling and keeping the plot easily interpretable. It can be seen that the plotted error bars are large around the edges of the regions exhibiting the highest flux. This is likely due to the fact that the chosen annulus is still wide enough that our calculations are likely to capture areas with little or no signal in addition to areas with high excess nebulosity, thus increasing the standard error. It should be noted that the median across all angles has been subtracted from each of the azimuthal profiles in Figure \ref{fig:angleprof} to better highlight variation in the data -- as such, some regions of the brighter azimuthal profiles might seem overly negative, but this is simply because they have a larger median to subtract, and is not indicative of significant levels of oversubtraction. For both the radial and azimuthal profiles we elect to present the median surface brightness as this is less sensitive to biasing from a small number of unusually bright/dim pixels than the mean.

In addition to the radial and azimuthal profiles of our three targets in Figures~\ref{fig:radprof} and \ref{fig:angleprof}, we also include the profile of the classical reduction of reference star 3 using reference star 2 as presented in the final row of Figure \ref{fig:finalplots} as an example of a null detection. It should be noted that the large dip in flux for the reference reduction exhibited around $260^{\circ}$ in Figure \ref{fig:angleprof} is not due to any astrophysical phenomenon but rather results from slight telescope jitter that was not covered by our enlarged BAR5 mask. This can be seen in Figure \ref{fig:finalplots} as a thin line of blue pixels at the South edge of the BAR5 mask, and will not have adversely affected our target reductions due to the use of multiple telescope orientations.

We now examine in greater detail each of our three target reductions in turn, with the use of the images presented in Figure \ref{fig:finalfinalplot} and the radial and azimuthal profiles presented in Figures \ref{fig:radprof} and \ref{fig:angleprof}.

\subsection{J16090075--1908526}
\label{subsec:t1}

The only potential signal present in the final image for this target is very close to the mask. However, this very quickly tends to the level of background noise, as evidenced by the comparison of the radial profiles of this reduction to that of the diskless reference reduction, which are almost indistinguishable apart from the unreliable inner regions. \citet{barenfeld2017} reports that this system has an inclination of $56^{+5}_{-5}$ degrees to the line of sight in observations of the $0.88$~mm continuum (see Table \ref{tab:observations}), and as such the fact that the azimuthal profile of this target shows little or no directionality is also evidential of a non-detection. As such, we take this result to be a non-detection of this target's disk. This may be due to the fact that the extent of the disk when observed at other wavelengths is almost entirely obscured by our mask (see Section \ref{subsec:comparisons} for further details). It should also be noted that, as described in Section \ref{datareduction} above, no additional structure could have been revealed using a smaller mask due to the unreliability of the residuals in these inner regions.

\subsection{J16142029--1906481}
\label{subsec:t2}

The apparent signal in this image is concentrated West of the star, as indicated by the arrow in Figure \ref{fig:finalfinalplot}. This can also be seen in the azimuthal profile in Figure \ref{fig:angleprof}, with a peak around $270^{\circ}$ to $300^{\circ}$ that is consistently two to three standard errors higher than the reference reduction. The directionality of the excess nebulosity makes the radial profile less pronounced, as the bright regions are averaged out by a lack of signal at other angles within the same annulus. Nonetheless, there is a plateau in the radial surface brightness profile around $1\farcs2$ to $1\farcs5$ (or projected distances of ${\sim}160$--$200$~au) that approximately corresponds to the region of excess nebulosity highlighted in Figure \ref{fig:finalfinalplot}. The steps detailed above to eliminate false positive detections leave us satisfied that this is indeed a true disk signal. Additionally, although we consider the classical reduction the most reliable, the azimuthal asymmetries that indicate disk structure are present in the reductions for both NMF and KLIP, increasing our confidence that the structures we observe are real and physical. Given the tentative nature of our detection, none of Figures \ref{fig:finalfinalplot}, \ref{fig:radprof} and \ref{fig:angleprof} are individually sufficient evidence of a disk detection, and it is only when taken holistically that we can see that a disk is indeed present.

\subsection{J16123916--1859284}

This target exhibits our highest-confidence detection of a protoplanetary disk. The flux of the disk is greater than that of any other reduction out to separations of 2\farcs~(or a projected distance of ${\sim}280~\textrm{au}$), consistently above the diskless reference reduction profile (at least 3 sigma above within the region of interest between the two vertical dashed lines). As seen in Figure \ref{fig:finalplots}, this signal is also highly invariant under method of reduction, increasing confidence that the signal observed is physical and in no way method-dependent. As with J16142029--1906481, this signal has an angular dependency, exhibiting the bulk of its flux due West of the star. This bump in flux is visible in Figure \ref{fig:angleprof}, again sitting at least three sigma above the reference profile. This signal appears to be in the same position angle as the disk for J16142029--1906481, which might in some cases be an indicator that this is merely a PSF-residual artifact appearing in the same position in both images. However, in our case this is purely coincidental, as the two disks were originally observed at very different orientations, as can be seen by the yellow arrows in Figure \ref{fig:finalfinalplot} indicating the North positions of the original observations in the detector frame. As these arrows show, these features appear on opposite sides of the unrotated PSF and thus cannot be the same PSF artefact exhibited in multiple images. The false-positive analysis described above also applies here, again increasing confidence that this is a true disk signal.

\subsection{Comparisons with other studies}
\label{subsec:comparisons}

\subsubsection{Scattered light observations}

Since our observations were made, two of our targets have been observed in the near-infrared as part of the DARTTS-S survey \citep{garufi2020}. The results for J16090075--1908526 and J16142029--1906481 presented in \citet{garufi2020} indicate that both of these targets do indeed host disks visible in scattered light. The radial extent of these disks is such that all of the observed structure shown in \citet{garufi2020} is obscured by our mask, as can be seen by the overplotted DARTTS-S disks in the corresponding images in Figure \ref{fig:finalfinalplot}. Previous work has shown that STIS is able to observe extended structure not visible at longer wavelengths (e.g. the halo reported around HR 4796A in \citealt{schneider2018} that goes undetected in \citealt{milli2019}; also the extended halo around HD 191089 visible using STIS but not with the Gemini Planet Imager detailed in \citealt{ren2019}). Our detection of J16142029--1906481 illustrates this effect, as our image exhibits significant flux on the West side of the star and little or no flux on the East side, similar to the images presented in \citet{garufi2020} and implying that what we are observing is a continuation of the structure observed as part of DARTTS-S. 

\citet{mawet2017} also notes the difference of disk surface brightness when comparing STIS and SPHERE observations of HD 141569 A, and hypothesises that STIS is able to probe smaller dust grains than SPHERE, and that these smaller grains in their disk have been swept out to large radii as a result of stellar radiation. Deeper follow-up imaging of our disks is required to be able to confirm the structure observed and to better probe the true nature of the observed dust grains. Comparisons between STIS and SPHERE are not possible with our final science image of J16090075--1908526, as we have observed an azimuthally symmetric non-detection, whereas the DARTTS-S disk has excess nebulosity to the West of the star. However, this non-detection can still be used to constrain the radial extent of the small dust grains in the disk to within the projected radius of our mask (${\lesssim}80$~au).

\subsubsection{Submillimetre observations}

We can compare our observations with the \citet{barenfeld2016} ALMA data that initially highlighted these three targets for our follow-up observations in scattered light. \citet{barenfeld2017} reports the results of fitting a surface density model to the \citet{barenfeld2016} $0.88$~mm continuum and $^{12}$CO $J = 3-2$ data to characterise the spatial extent of the observed disks in both regimes. The results they obtained are presented in Table \ref{tab:observations} and overplotted in Figure \ref{fig:finalfinalplot} -- while the detections presented herein are too tentative to perform analogous model fitting, a cursory comparison between the \citet{barenfeld2017} results and our own images can still be performed.

The ALMA $0.88$~mm continuum and $^{12}$CO $J = 3-2$ disk radii for each of our detected disks are, as with the SPHERE data, smaller than the central region of the mask used in this work, but we can still compare other features. For example, the reported position angle of the semi-major axis $\theta_{\textrm{CO}}$ for the J16123916--1859284 disk is approximately aligned with what we observe in Figure \ref{fig:angleprof}, with $\theta_{\textrm{dust}}$ at an offset of ${\sim}45^{\circ}$ from our scattered light disk. This can be seen using the ellipses in Figure \ref{fig:finalfinalplot} and provides tentative evidence that the extended small dust grains trace the gas distribution in the disk, as would be expected for dust grains entrained in the gas. 

The position angle data for both types of detected ALMA emission imply that the excess nebulosity for J16142029--1906481 is approximately parallel to the semi-minor axis of the \citet{barenfeld2017} disk, although there are considerable uncertainties associated with this particular fit to the data. Without higher signal-to-noise data and detailed disk modelling it is difficult to know if our observations agree or disagree with these ALMA data, and this would be another good focus for future work.

Whilst we detect no significant structure around J16090075–1908526, \citet{barenfeld2016} finds this disk to be the most extended in CO and in 0.88mm continuum emission of those observed as part of our study. This indicates that the micron-sized dust grains are poorly coupled to both the CO gas and larger mm-size dust grains at these extended radii, in agreement with results reported in \citet{villenave2019} and \citet{rich2021} and contrasting with the relationship we observe for our detected disks. Again, further characterisation via deeper imaging and disk modelling could potentially help to resolve these seemingly contradictory sets of observations. This further modelling could also give insight into the masses of the small dust grains observed in the disk, and the properties of these dust grains, both of which are beyond the scope of the marginal detections presented in this paper.



\section{Conclusion}

We report \textit{HST}/STIS observations of three systems of M-type stars in Upper Sco known to host protoplanetary disks visible at ALMA wavelengths. We have experimented with three different methods of Reference Differential Imaging, and found that a classical reduction scaled by minimising the standard deviation in the PSF diffraction spikes balances reliability, undersubtraction and oversubtraction best of these three methods.
In the classically reduced images of our three systems, we tentatively detect disks around 2MASS J16142029--1906481 and 2MASS J16123916--1859284. We fail to detect a disk around our third target, 2MASS J16090075--1908526.

Both of our detected disks exhibit structure out to projected distances of ${\gtrsim}200~\textrm{au}$, further from the star than any structure previously detected for either disk. By comparison with the radial extent of available SPHERE data for the disk around J16142029--1906481, we have shown that visible-wavelength STIS observations are better able to probe dust grains out to greater radii than other instruments, in agreement with previous work. Our work adds to the relatively small sample of images of resolved disks around young M-type stars, and highlights the necessity of further characterising the dust distributions around these disks, either by making deeper observations or by using detailed disk modelling, in order to more precisely characterise these protoplanetary disks.

\section*{Data Availability}

The data underlying this article are available in their raw form via the \href{https://archive.stsci.edu/hst}{HST MAST archive} under programs \href{https://www.stsci.edu/hst/phase2-public/15176.pdf}{GO 15176} and \href{https://www.stsci.edu/hst/phase2-public/15497.pdf}{GO 15497}. Processed data are available from the author on request.

\section*{Acknowledgements}

We thank John H.~Debes for providing the algorithmic mask in \citet{debes2017}. This research has made extensive use of {\tt numpy} \citep{numpy}, {\tt scipy} \citep{scipy} and {\tt matplotlib} \citep{matplotlib}. The results of this paper are based on observations made with the NASA/ESA \textit{Hubble Space Telescope}. We thank support from \href{https://www.stsci.edu/hst/phase2-public/15176.pdf}{GO 15176} and \href{https://www.stsci.edu/hst/phase2-public/15497.pdf}{GO 15497} provided by NASA through a grant from STScI under NASA contract NAS 5-26555. JMC acknowledges support from the National Aeronautics and Space Administration under grant No. 15XRP15\_20140 issued through the Exoplanets Research Program.

\bibliographystyle{mnras}
\bibliography{bib}

\bsp	
\label{lastpage}
\end{document}